\documentclass[11pt]{article}
\usepackage{graphicx}

\begin{document}

\def\xslash#1{{\rlap{$#1$}/}}
\def \p {\partial}
\def \dd {\psi_{u\bar dg}}
\def \ddp {\psi_{u\bar dgg}}
\def \pq {\psi_{u\bar d\bar uu}}
\def \jpsi {J/\psi}
\def \psip {\psi^\prime}
\def \to {\rightarrow}
\def\bfsig{\mbox{\boldmath$\sigma$}}
\def\bfeps{\mbox{\boldmath$\epsilon$}}
\def\DT{\mbox{\boldmath$\Delta_T $}}
\def\xit{\mbox{\boldmath$\xi_\perp $}}
\def \jpsi {J/\psi}
\def\bfej{\mbox{\boldmath$\varepsilon$}}
\def \t {\tilde}
\def\epn {\varepsilon}
\def \up {\uparrow}
\def \dn {\downarrow}
\def \da {\dagger}
\def \pn3 {\phi_{u\bar d g}}

\def \p4n {\phi_{u\bar d gg}}

\def \bx {\bar x}
\def \by {\bar y}

\begin{center}
{\Large\bf Near Threshold Enhancement of $p\bar p$ System and $p\bar p$ Elastic Scattering}
\vskip 10mm
G.Y. Chen$^1$, H.R. Dong$^2$ and J.P. Ma$^{2,3}$    \\
{\small {\it $^1$ Department of Physics, Peking University,
Beijing 100871, China }} \\
{\small {\it $^2$ Institute of Theoretical Physics, Academia Sinica,
Beijing 100190, China }} \\
{\small {\it $^3$ Theoretical Physics Center for Science Facilities, Academia Sinica,
Beijing 100049, China }} \\
\end{center}
\begin{abstract}
The observed enhancement of $p\bar p$-production near the threshold in radiative decays of $J/\psi$ and
$e^+e^-$-annihilations can be explained with final state interactions among the produced
$N\bar N$ system, where the enhancement is essentially determined by $N\bar N$ elastic scattering amplitudes.
We propose to use an effective theory for interactions in a $N\bar N$ system near its threshold.
The effective theory is similar
to the well-known one for interactions in a $NN$ system but with distinctions. It is interesting to note that
in the effective theory some corrections to scattering amplitudes at tree-level can systematically be
summed into a simple form. These corrections are from rescattering processes. With these corrected
amplitudes we are able to describe the enhancement near the threshold in radiative decays of $J/\psi$ and
$e^+e^-$-annihilations, and the $p\bar p$ elastic scattering near the threshold.
\end{abstract}

\par\vskip20pt
\par
It has been observed the enhancement of  the $p\bar p$-production
near the threshold in various experiments. The enhancement has been
observed near the threshold of the $p\bar p$ system in the decay
$J/\psi \to \gamma  p \bar p$ by BES\cite{BES1} and in the decay of
$B^+\to K^+ p\bar p$ and $\bar B^0 \to D^0 p\bar p$ by
Belle\cite{Bell}. The enhancement also has appeared in the process $e^+
e^- \to p\bar p$ measured by Babar\cite{Bar1}. The enhancement of other baryonic
system has been also seen\cite{BES2,Bar2}. In this work we focus on
the enhancement of $p\bar p$ system.
\par
Because BES was the first
to publish the result about the enhancement with rather high statistical accuracy
and precise information about the spectrum,
many explanations for the observed enhancement at BES exist. A class of explanations is that
the enhancement is interpreted as the existence of a baryonium bound state\cite{Tbes1} or a glueball
below the threshold\cite{BesG}.
Another class of explanations is to take the effect of final state interactions into account. There are different
ways to take final state interactions into account. One can use a complex $S$-wave $p\bar p$ scattering length\cite{Tbes2}
or use a $K$-matrix formalism to include one pion exchange\cite{ZC}.
A more realistic way is by using potential models of $N\bar N$ interactions\cite{Tbes3,Tbes4}.
The observed enhancement in $e^+e^- \to p\bar p$ has also motivated theoretical
studies\cite{Tbes4,HML,Tbell,Tbar1,Tbar2,Tbar3}. It is interesting to note that by taking final state interactions
into account through potential models of $N\bar N$ interactions, the enhancement in $J/\psi \to \gamma p\bar p$ and
$e^+e^- \to p\bar p$ can be explained simultaneously\cite{Tbes4,Tbar2}. However, these models are in general complicated
and contain several or more parameters which need to be fixed. It should be noted that
with final state interactions the enhancement is predicted
with the energy dependence of $p\bar p$-elastic scattering amplitudes, where one employs
Watson-Migdal approximation\cite{WM}.
\par
A rather simple approach with final state interactions for explaining the enhancement has been give in \cite{CDM}.
The $p\bar p$-elastic scattering amplitude there is determined through the rescattering mechanism similar
to Watson-Migdal approximation, where one resums the multi-pion exchange or multi-rescattering of a $N\bar N$ system.
Because the coupling of $\pi NN$
is well-known, the amplitude is completely fixed in this approach.
It has been shown in \cite{CDM} that the observed enhancement
in the $J/\psi\to \gamma p\bar p$ and $e^+e^- \to p\bar p$ can be well explained. But, with the fixed
amplitude one can not explain the $p\bar p$ elastic scattering near the threshold. In this work
we make an attempt in the framework
of effective field theories to give an unified explanation  for the enhancement in the $J/\psi$ decay
and $e^+e^-$ annihilation,
and for the $p\bar p$ elastic scattering near the threshold.
\par
An effective theory of $N\bar N$ interactions can be developed in analogy to the effective theory
of $NN$ interactions. The effective theory of $NN$ interactions has been proposed in\cite{Wein,vKK,EFTN}
and studied extensively in \cite{vKK,EFTN,EFTN1}. With the effective theory the experimental data
of $NN$ scattering of low partial waves near the threshold can be well described.
In constructing such effective theories one makes
an power expansion in the momentum near the threshold. The coefficients in the expansion
characterize the properties of the $NN$- or $N\bar N$ system, like scattering length $a_0$, interaction
range $r_0$, etc..
\par
There are distinct differences between the effective theory of $N\bar N$ interactions
and that of $NN$ interactions.
As for an effective field theory, one does not only need to construct the effective Lagrangian
which gives tree-level amplitudes directly,
but also one should be able to estimate relative importance of higher order corrections
and to systematically calculate these corrections. In other word, one needs a power counting
for the estimation of loop corrections. If the interaction of a system is characterized by a momentum scale $\Lambda$
and the quantities like $a_0$, $r_0$,..... have the natural size, i.e., $a_0 \sim \Lambda^{-1}$,
a simple power counting like that for the well-known chiral perturbation theory can be used. However, for $NN$ systems
it is not the case. It is well known from experiment that $NN$-systems have large scattering lengthes. This fact makes
the simple power counting invalid. An important idea has been suggested in \cite{EFTN}, where one can use
the power divergence subtraction scheme
instead of the minimal subtraction scheme which is commonly used, implemented to the effective theory.
With the power divergence subtraction scheme a power counting can be established. This makes
the effective theory well defined for $NN$ system with large scattering lengthes.
In the case of $N\bar N$ systems, the scattering lengthes,
according LEAR experiment\cite{Tbes2} and model results\cite{Tbes3}, are around $1{\rm fm}$ or smaller.
They are much smaller than those of $NN$ systems. Therefore one can use the minimal subtraction scheme and
hence the simple power counting
for the effective theory of $N\bar N$ systems.
\par
Another difference is that a $N\bar N$ system
can be annihilated in to a multiple pion system and the pions can be real, while a $NN$ system
can not be annihilated. The annihilation of a $N\bar N$ system into virtual- or real pions results in that
the dispersive- and absorptive part of the $N\bar N$ scattering amplitudes are of the same importance.
In order to incorporate this fact some coupling constants in the effective theory of a $N\bar N$ system
are complex numbers. In the effective theory of a $NN$ system the coupling constants are real.
In this work we will first discuss the effective theory of a $N\bar N$ system and the $NN$ elastic scattering.
Then we study the rescattering mechanism for the enhancement in the decay
$J/\psi \to \gamma p \bar p$ and the annihilation $e^+ e^- \to p\bar p$. We will show that
with our effective theory approach the enhancement in the decay
$J/\psi \to \gamma p \bar p$ and the annihilation $e^+ e^- \to p\bar p$
and the $p\bar p$ elastic scattering near the threshold can be well described.
\par
We consider the $N\bar N$ scattering near the threshold:
\begin{equation}
 N(\vec p,s_1) + \bar N(-\vec p,s_2) \to N(\vec k,s_1') + \bar N(-\vec k,s_2'), \ \ \
 {\vert \vec p \vert}=p= \beta {\sqrt{m^2 + p^2}},
\end{equation}
where $\vec p$ and $\vec k$ are three-momenta. The spins are denoted with $s$'s.
$\beta$ is the velocity. Near the threshold, the momentum $p$ or
$\beta$ approaches to zero. We are interested in the momentum region
$p\leq m_\pi$.  An effective Lagrangian can be obtained by an
expansion in $p$ or $\beta$. For this it is natural to use
nonrelativistic fields to describe the nucleon $N$. The
nonrelativistic fields are given as
\begin{equation}
\psi = \left ( \begin{array}{cc} \psi_p \\ \psi_n \end{array} \right ),\ \ \ \
\chi = \left ( \begin{array}{cc} \chi_p \\ \chi_n \end{array} \right ). \ \ \ \
\end{equation}
The two-component field $\psi_p (\chi^\dagger_p)$ annihilates a proton(an anti-proton) and
the field $\psi^\dagger_p (\chi_p)$ creates a proton(an anti-proton).
We denote the Pauli matrix acting in the $SU(2)$ isospin space as $\tau^i(i=1,2,3)$. The
$\pi$ fields are given as $\pi = \tau^i \pi^i$ with $\pi^3 =\pi^0$.
At leading order the interacting
part can be written as:
\begin{eqnarray}
\delta{\mathcal L} &=& \frac{c_0}{4} \psi^\dagger \chi \chi^\dagger \psi
   + \frac{c_1}{4}\psi^\dagger \tau^i \chi \chi^\dagger \tau^i \psi
   +\frac{d_0}{4} \psi^\dagger \sigma^i \chi \chi^\dagger \sigma^i \psi
   + \frac{d_1}{4}\psi^\dagger \tau^i \sigma^j \chi \chi^\dagger \sigma^j \tau^i \psi
\nonumber\\
   && \ \ + \frac{g_A}{2 F_\pi} \psi^\dagger \vec\sigma\cdot \left ( \vec\partial \pi \right ) \psi
          + \frac{g_A}{2 F_\pi} \chi^\dagger \vec\sigma\cdot \left ( \vec\partial \pi \right ) \chi
           + {\mathcal O}(p^2),
\end{eqnarray}
with $g_A \approx 1.25$ and $F_\pi \approx 93{\rm MeV}$. $c_I$ with $I=0,1$ is the coupling constant
in the $^1 S_0$ channel, while $d_I$ with $I=0,1$ is the coupling constant
in the $^3 S_1$ channel.
As discussed before,
these coupling constants are in general complex because a $N\bar N$ can annihilate
into $\pi$'s. One should keep in mind that the complex coupling constants here do not mean the violation
of time-reversal symmetry. The complex coupling constants can be understood as the following: One can imagine
that the effective theory is obtained from a perturbative matching of a more fundamental theory. In the more
fundamental theory with the time-reversal symmetry amplitudes at tree-level are real, but
they receive imaginary parts beyond tree-level because absorptive parts are nonzero at one- or more loop level.
The imaginary parts of the coupling constants are from these absorptive parts in the matching.
\par
To clearly discuss the mentioned simple power counting for the above effective theory we first ignore
the interactions of pion exchanges. Then in this case, the scattering amplitude at tree level is expanded
in $p$, the leading orders are determined by the contact interactions given in Eq.(3)
and are at ${\mathcal O} (p^0) $ if we take $c_{0,1}$ and $d_{0,1}$ as constants.
The tree-level contributions at higher orders of $p$ starting at ${\mathcal O} (p^2) $ are given by operators with derivatives
in the effective theory. Therefore, at tree-level the amplitude is simply expanded in power of $p$
and the power of $p$ of each term is determined by the corresponding contact terms in the effective theory.
However, this can be changed if we take loop-effects into account, i.e., the effects of scale-dependence
of coupling constants. E.g., in the effective theory for a $NN$ system of large scattering lengthes
the coupling constants corresponding to $c_{0,1}$ and $d_{0,1}$ should be taken
as at order of $p^{-1}$ after including loop effects with power subtraction scheme,
where one reasonably takes the renormalization scale as $\mu\sim p$. In this way,
a consistent power counting is established for a $NN$ system with large scattering lengthes\cite{EFTN}.
\par
As discuss before, it is expected that quantities characterizing interactions of
a $N\bar N$ system have the nature size $\Lambda$. With this expectation the coefficients $c_{0,1}$ and $d_{0,1}$ in
the effective theory scale like $\Lambda^{-2}$ or $(m\Lambda)^{-1}$ with $m$ as the nucleon mass.
They are at order of ${\mathcal O} (p^0)$ with
the minimal subtraction scheme in dimensional regularization as shown in \cite{EFTN}.
The scale-dependence of these coupling constants are suppressed by certain power of $p$.
The loop contributions formed only by the contact interactions can be then estimated as the following: Each loop
contributes a factor $(mp/4\pi)$. Hence, the leading contribution is at $p^0$ and comes from
$c_{0,1}$ and $d_{0,1}$ at tree-level. The next to-leading order is at $p^{1}$ and comes from
$c_{0,1}$ and $d_{0,1}$ at one-loop level. The contribution at $p^2$ comes from
$c_{0,1}$ and $d_{0,1}$ at two-loop level and from dimension-8 operators in $\delta {\mathcal L}$
at tree-level, etc. In the above we discuss the simple power counting without interactions with pions.
Since we consider the momentum region of $p\leq m_\pi$,
the interaction with $\pi$ is taken at the order of ${\mathcal O}(p^0)$.
In Eq.(3) the leading interactions are given explicitly.
At higher order of $p$ operators with derivatives
and operators for emission of more than one pion will appear.
We notice that the power counting of loop-contributions through exchange of pions is slightly different
because some loop contributions can be more strongly suppressed than the estimated
by the simple power counting.
\par
With the interactions given in the above and also for our purpose,
it is convenient to work with partial waves of the scattering. The
scattering amplitude of Eq.(1), denoted as ${\mathcal T}_I(\vec
p,\vec k,s_1,s_2,s_1',s_2')$ with the isospin $I=0,1$, can be decomposed
with CG coefficients and harmonic oscillators into partial waves ${\mathcal T}_{[j\ell\ell' s,I]}(E)$:
\begin{eqnarray}
 {\mathcal T}_I(\vec p,\vec k,s_1,s_2,s_1',s_2')
  &=& \sum
     Y^*_{\ell\ell_3}(\vec p/p) Y_{\ell'\ell'_3}(\vec k/k)
   (1/2,s_1;1/2,s_2\vert s,s_3) ( 1/2, s'_1;1/2,s'_2 \vert s',s'_3)
\nonumber\\
  && \cdot (\ell,\ell_3,s,s_3\vert j,j_3) (\ell',\ell'_3,s',s'_3\vert j',j'_3)
  \delta_{jj'} \delta_{j_3j_3'}
    \delta_{ss'}  \ {\mathcal T}_{[j\ell\ell' s,I]}(E).
\end{eqnarray}
In the above the repeated indexes are summed and $E$ is the total kinetic energy $E=p^2/m$
of the system. With the effective Lagrangian
it is clear that the term with $c_{I}$ will only contribute to ${\mathcal T}_{[0000,I]}(E)$ with $I=0,1$, respectively.
\par
With the effective theory it is straightforward to work out the scattering amplitude and various
partial waves at tree-level to compare with experimental results of $p\bar p$-scattering and enhancement
near the threshold. In the comparison we perform a combined fit to fit the cross-section of $p\bar p$-scattering
and the enhancement in $J/\psi\to \gamma\bar p$ and $e^+e^-\to p\bar p$, where the coupling constants $c_{0,1}$
and $d_{0,1}$ are taken as free parameters.
However, as we will mention and show later, the theoretical results at tree-level
can only fit the experimental data in a small region with $E\leq 20$MeV. Although the $\chi^2/d.o.f.$
of the fit is close to $1$, but the coupling constants can only be determined
with the error from $40\%$ to $100\%$ or even more.
In this work we will improve the situation by adding some corrections beyond tree-level.
In the perturbation expansion of the effective theory some corrections appearing in $n$-loop level can be summed
into a compact form. We will add these corrections to our tree-level results at tree-level
and show that the improvement is significant. In the below we will study these corrections. This also
illustrates some aspects of the effective theory.

\par


\begin{figure}[hbt]
\begin{center}
\includegraphics[width=12cm]{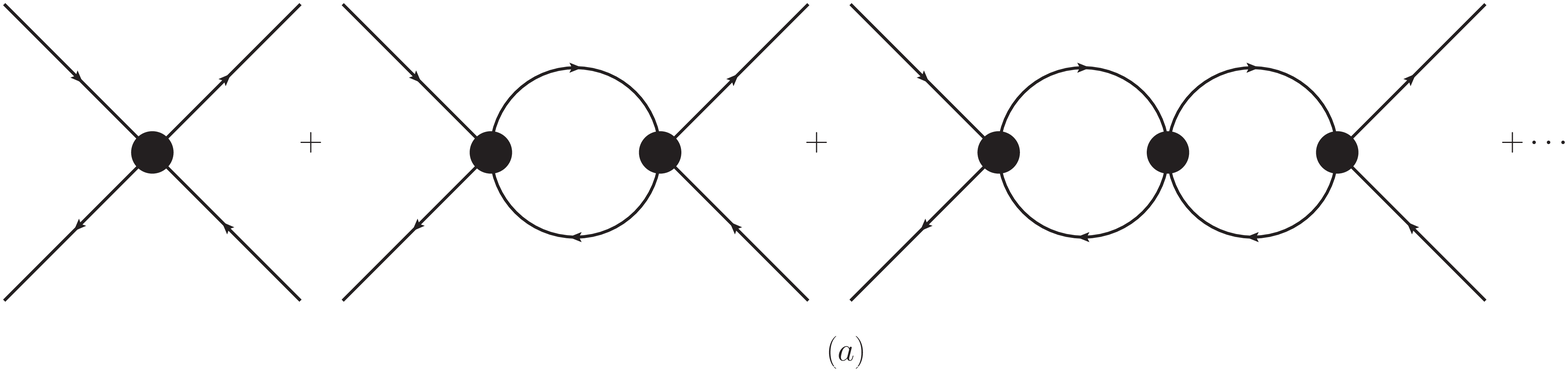}
\end{center}
\begin{center}
\includegraphics[width=12cm]{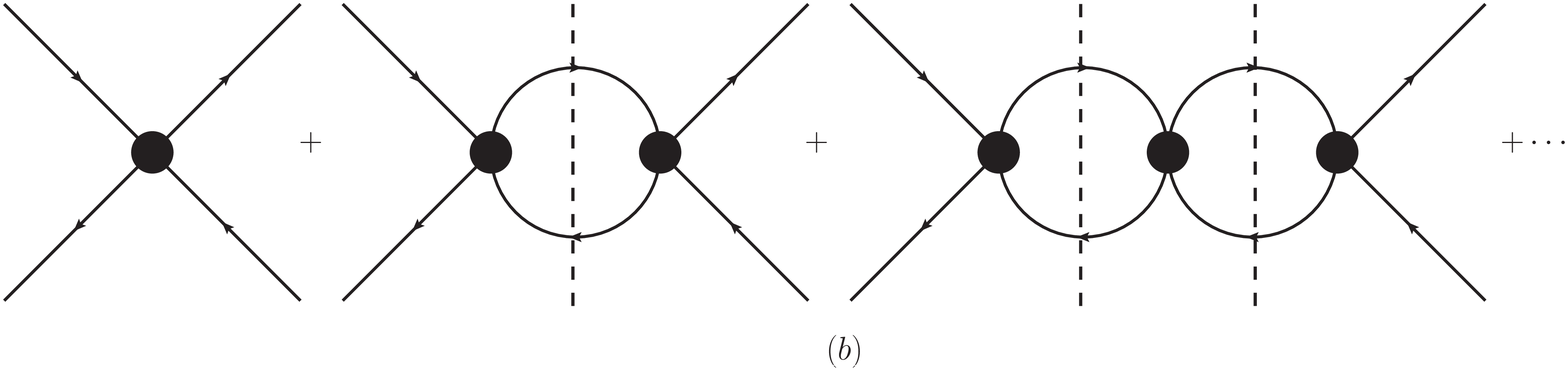}
\end{center}
\caption{The loop diagrams with the vertex of $c_0$. The black dots denote vertices of the contact term.}
\label{Feynman-dg2}
\end{figure}
\par
We focus at moment on the contribution only from $c_0$ by ignoring the contribution from $\pi$-exchange.
It is straightforward to obtain the contributions from tree-, one-loop,
etc. as given from the diagrams in Fig.1a. The tree-level amplitude is just $4\pi c_0$.
The one-loop contribution is proportional to the loop integral regularized in $d$-dimension\cite{EFTN}:
\begin{eqnarray}
I_0 &=& i \mu^{4-d}\int \frac{d^d q}{(2\pi)^d} \left [ q^0 -\frac{\vert\vec q \vert ^2}{2m} + i\varepsilon \right ]^{-1}
  \left [ E-q^0 - \frac{\vert \vec q \vert ^2}{2m} + i\varepsilon \right ]^{-1}
\nonumber\\
    &=& -m \left (-mE-i\varepsilon\right  )^{\frac{d-3}{2} } \Gamma \left(\frac{3-d}{2}\right ) \frac{\mu^{4-d}}{(4\pi)^{(d-1)/2}}.
\end{eqnarray}
An interesting observation can be made for the above result. The integral is finite with $d=4$. It has a pole at $d=3$
corresponding to a power divergence of the integral. The power divergence subtraction scheme is to subtract
from the above contribution
the pole contribution at $d=3$. The subtraction introduces the renormalization scale $\mu$, hence
a $\mu$-dependence of the coupling constant. With this scheme one can show that for $NN$-interactions with large scattering lengthes
a consistent power counting can be established\cite{EFTN} by setting $\mu\sim p \sim m_\pi$.
As discussed before, for $N\bar N$-interactions scattering lengthes are much smaller
than those from $NN$-interactions. Therefore we can employ the minimal subtraction scheme in which one subtracts
the pole terms at $d=4$. Because the above one-loop integral is finite at $d=4$, no subtraction is needed. With $d=4$
we have
$4\pi I_0 = -im p=-im^2\beta$. Inspecting the contribution from $n$-loop, one will find that the contribution is proportional to $I_0^n$.
In fact the sum of $n$-loop contributions forms a sum of a geometric series. The sum can easily be performed. We have then
the exact the amplitude without $\pi$-exchange as:
\begin{equation}
{\mathcal T}_{[0000,0]}(E)\biggr \vert_{c_0} = \frac {4\pi c_0} { 1 -i\frac {m^2\beta }{4\pi} c_0}.
\end{equation}
Expanding the above expression in $c_0$ one can identify that the term with $c_0^{1+n}$ comes
from the $n$-loop diagram. Through the expansion one also sees that
each loop brings a suppression factor $p$ or $\beta$ as indicated by the discussed simple power counting.
The tree-level amplitudes obtained from the effective Lagrangian in Eq.(3) are at order of ${\mathcal O}(p^0)$.
The corrections to them start at order of ${\mathcal O}(p)$. By considering corrections from higher orders in $p$,
the coupling constants in our effective theory will be generally depend on the renormalization scale $\mu$.
With the power counting the $\mu$-dependence is suppressed at least by ${\mathcal O}(p)$.
\par

\begin{figure}[hbt]
\begin{center}
\includegraphics[width=12cm]{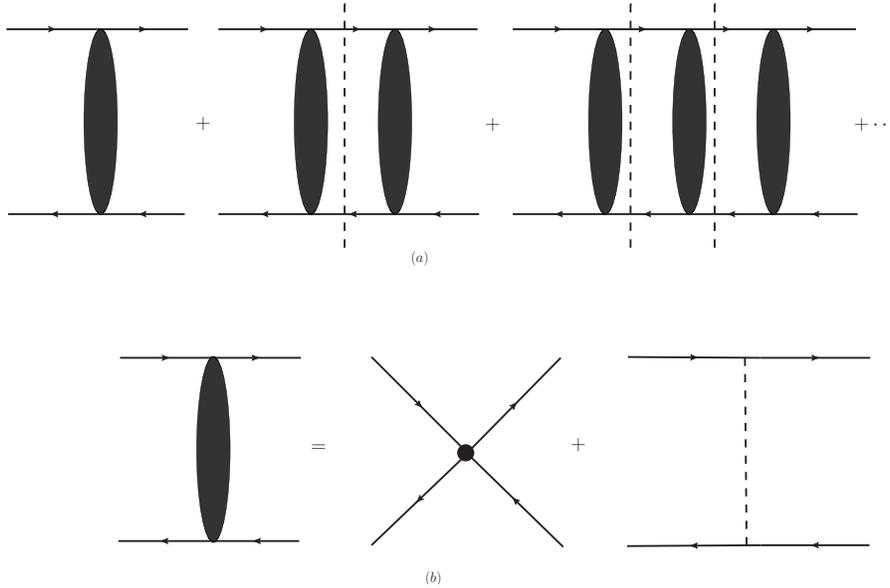}
\end{center}
\caption{Fig.2a represents the sum of rescattering amplitudes, where the nucleon- and antinucleon lines going
through the cuts are on-shell. The narrow and long bubble represents the tree-level amplitude with
the contact interactions and pion-exchange as indicated by Fig.2b.
} \label{Feynman-dg2}
\end{figure}

\par
Observing the above result one can realize that the loops in Fig.1a calculated with the minimal subtraction scheme
are the same in Fig.1b calculated only by taking the absorptive part of the loop integral $I_0$, i.e., calculated
by putting the $N\bar N$ in the loop on-shell indicated by the cuts in Fig.1b. Now we consider to include the contributions
from $\pi$-exchanges. The tree-level contribution can be represented by the first diagram of Fig.2a where the bubble contains
the vertex of $c_0$ and one-$\pi$-exchange as indicated in Fig.2b.
The contribution of Fig.2b is straightforward
to obtain:
\begin{equation}
{\mathcal T}_{[0000,0]}^{(0)}(E) = 4\pi \left [ c_0 -\frac{3}{2} g_0^2 \left (2 - y L_y \right )\right ],
\end{equation}
with the notations:
\begin{equation}
  g_0^2 = \left ( \frac {g_A}{2 F_\pi }\right )^2, \ \ \ \  y= \frac{m_\pi^2}{2 p^2}, \ \ \  L_y =\ln \left ( 1 +\frac{2}{y}\right).
\end{equation}
It is easy to find that the sum of bubble diagrams with cuts in Fig.2a can be performed. It is a geometric series. We have then the summed
amplitude:
\begin{equation}
{\mathcal T}_{[0000,0]}(E) = {\mathcal T}_{[0000,0]}^{(0)}(E)
\left [ 1 -i\frac{m^2\beta}{(4\pi)^2} {\mathcal T}_{[0000,0]}^{(0)}(E) \right]^{-1} + \cdots,
\end{equation}
where $\cdots$ stand for other corrections, like the dispersive part of one-loop contributions formed
with $\pi$-exchanges. In the work we will neglect these corrections for our predictions. These
corrections can be systematically studied. We will come back to the dispersive part in a future work,
where a complete one-loop calculation will be performed.
Physically the interpretation of Fig.2a  and the amplitude in Eq.(9)
is the following: The $N\bar N$ undergoes a multiple scattering process
$N\bar N \to N\bar N \to \cdots \to N\bar N$. Each scattering is due to the vertex with $c_0$ or exchange of one pion.
Each pair of $N\bar N$ is on-shell in Fig.2.
We will call amplitudes for such a multiple scattering process as rescattering amplitudes.
Similarly
one can work out the tree-level and rescattering amplitude with $I=1$:
\begin{eqnarray}
{\mathcal T}_{[0000,1]}^{(0)}(E) &=& 4\pi \left [ c_1 +\frac{1}{2} g_0^2 \left (2 - y L_y \right )\right ],
\nonumber\\
{\mathcal T}_{[0000,1]}(E) &=& {\mathcal T}_{[0000,1]}^{(0)}(E)
\left [ 1 -i\frac{m^2\beta}{(4\pi)^2} {\mathcal T}_{[0000,1]}^{(0)}(E) \right]^{-1} + \cdots.
\end{eqnarray}
\par
Now we turn to the amplitudes with $j=1$ and $s=1$. In this case it is little complicated because
the $\pi NN$-interaction mixes amplitudes with difference $\ell$. The difference caused by exchanging one $\pi$
can only be $\pm 2$ . The summed
or rescattering amplitude has to be expressed in a matrix form.  We define the following matrix amplitude:
\begin{equation}
{\mathcal M}_{[I]}(E)  =   \left (\begin{array}{cc} {\mathcal T}_{[1001,I]}(E),
&  {\mathcal T}_{[1201,I]}(E)
 \\  {\mathcal T}_{[1021,I]}(E) , & {\mathcal T}_{[1221,I]}(E) \end{array} \right ).
\end{equation}
The tree-level results for ${\mathcal M}_{[0]}(E)$ reads:
\begin{equation}
{\mathcal M}_{[0]}^{(0)}(E)  =  \pi \left (\begin{array}{cc} 4 d_0 + 2g_0^2\left ( 2 -y L_y \right ),
&  \sqrt{2} g_0^2
  \left ( 2 -6y +(2y+3y^2) L_y \right )
 \\ \sqrt{2}g_0^2
  \left ( 2 -6y +(2y+3y^2) L_y \right )  , &  g_0^2 \left [ 2+ 6y -(4y+3y^2)L_y \right ] \end{array} \right ),
\end{equation}
for $I=1$ the matrix ${\mathcal M}_{[1]}^{(0)}(E)$ is obtained by replacing $d_0$ with $d_1$ and $g_0^2$ with $-g_0^2/3$.
The summed or rescattering amplitude matrix from Fig.2 can be found as:
\begin{equation}
{\mathcal M}_{[I]}(E)  ={\mathcal M}_{[I]}^{(0)}(E) \left [ I - i \frac{m^2\beta}{(4\pi)^2}{\mathcal M}_{[I]}^{(0)}(E) \right ]^{-1}.
\end{equation}
It should be noted that the contact terms in the Lagrangian are involved only in the above discussed amplitudes.
Later we will use these amplitudes and those of other partial waves at tree-level to compare with experimental results
of the cross section of $p\bar p$ scattering near the threshold.

\par
\begin{figure}[hbt]
\begin{center}
\includegraphics[width=12cm]{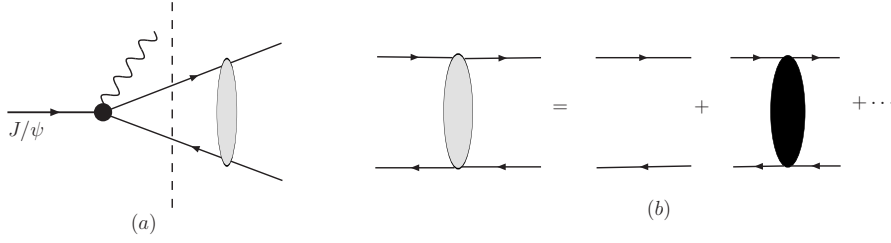}
\end{center}
\caption{The diagrammatic explanation of our approach. The dash line in Fig.3a is the cut.}
\label{Feynman-dg1}
\end{figure}
\par
Now we turn to the enhancement observed in $J/\psi \to \gamma p\bar p$ and $e^+e^-\to p\bar p$.
In \cite{CDM} we assumed that in these processes a $N\bar N$ system is produced first,
then the $N\bar N$ system undergoes a scattering before observed. This can be explained with Fig.3.
As illustrated in Fig.3, The production of a $p\bar p$ system
can be thought that at first step a $n\bar n$ system or a $p\bar p$ system is produced and then
the $N\bar N$ system through a rescattering is converted into the observed $p\bar p$ system.
We assume that the production of a $N\bar N$ system is a short distance process, i.e., the energy scale
characterizing the production is much large than $m_\pi$ and the momentum near the threshold.
Hence we can expand the vertex for the production in Fig.3a in the momentum $p$ or $\beta$.
At the leading order we can approximate
the black vertex for the production in Fig.1a with constant form factors.
We denote the form factors as ${\mathcal A}_{p\bar p,n\bar n}$, respectively.
At the leading order the $N\bar N$- hence the final $p\bar p$ system can only be
in the $^1S_0$ state.
Then the decay amplitude can be written as:
\begin{eqnarray}
{\mathcal T}_{J/\psi \to \gamma p\bar p} & \approx &  \varepsilon_{\mu\nu\sigma \rho}
  P^\mu_{J/\psi} k^\nu_\gamma \varepsilon_{J/\psi}^{\sigma} \varepsilon_\gamma^\rho
    \left\{ \frac{{\mathcal A}_{p\bar p}+ {\mathcal A}_{n\bar n} }{2}
     \left [ 1 + i\frac{ m^2 \beta }{(4 \pi)^2} {\mathcal T}_{[0000,0]}(E)  \right ]
     \right.
\nonumber\\
     && \left.  +\frac{{\mathcal A}_{p\bar p} - {\mathcal A}_{n\bar n} }{2}
     \left [ 1 + i\frac{ m^2 \beta }{(4 \pi)^2} {\mathcal T}_{[0000,1]}(E) \right ] \right\}
      + \cdots,
\end{eqnarray}
The $\cdots$ stand for those partial wave amplitudes with $\ell\geq 1$
and the corrections proportional to $\beta$.
With the approximation of the rescattering as discussed before, the $p\bar p$ amplitudes can be obtained from Eq.(9,10).
We have then for the decay amplitude near the threshold as
\begin{eqnarray}
{\mathcal T}_{J/\psi \to \gamma p\bar p} & \approx &  \varepsilon_{\mu\nu\sigma \rho}
  P^\mu_{J/\psi} k^\nu_\gamma \varepsilon_{J/\psi}^{\sigma} \varepsilon_\gamma^\rho
    \left \{ \frac{{\mathcal A}_{p\bar p}+ {\mathcal A}_{n\bar n} }{2}
      \left [ 1 - i\frac{m^2 \beta}{(4\pi)^2} {\mathcal T}_{[0000,0]}^{(0)}(E) \right]^{-1}
      \right.
\nonumber\\
     &&  \left. +\frac{{\mathcal A}_{p\bar p} - {\mathcal A}_{n\bar n} }{2}
     \left [ 1 - i\frac{m^2 \beta}{(4\pi)^2} {\mathcal T}_{[0000,1]}^{(0)}(E) \right]^{-1} \right \}.
\end{eqnarray}
Setting $c_{0,1}=0$ we recover our early results in \cite{CDM} where in the corresponding amplitude the factor $i$ should
be replaced with $-i$ because the absorptive part has been identified with a wrong sign.
The physical results in \cite{CDM} will be not changed with this replacement. In our approach
the intermediate state can only be $N\bar N$ systems according to our effective theory.
It is possible
to add contributions from mesons as intermediate states in some phenomenological models as shown in \cite{Xqian}.
\par
For the enhancement observed at Babar the form factors of proton are involved. These form factors are defined as:
\begin{eqnarray}
 \langle p(p_1) \bar p (p_2) \vert J^\mu \vert 0\rangle & = &  \bar u(p_1) \left [ \gamma^\mu F_1 (q^2)
      + i\frac{\sigma^{\mu\alpha}}{2 m_p} q_\alpha F_2(q^2) \right  ] v(p_2),
\end{eqnarray}
It should be noted that near the threshold only the combination of the two form factors is involved.
Near the threshold we have:
\begin{eqnarray}
\langle p(p_1) \bar p (p_2) \vert {\bf J } \vert 0\rangle = -2m_p \left ( F_1(q^2)+ F_2(q^2) \right )
 \xi^\dagger \bfsig \eta + {\mathcal O}(\beta^2).
\end{eqnarray}
Similarly, we introduce the mechanism as in Fig.3.
We characterize the vertex of the $N\bar N$ production at the first step with
constant form factors, denoted as $\tilde G_M^{(p)}$ and $\tilde G_M^{(n)}$.
Then the form factor with our approach is given as:
\begin{eqnarray}
F_1(q^2) + F_2(q^2) =
\frac{\tilde G_M^{(p)}+\tilde G_M^{(n)}}{2}
                       {\mathcal A}_{11}(0)
    +\frac{\tilde G_M^{(p)}-\tilde G_M^{(n)}}{2}
       {\mathcal A}_{11}(1)
\end{eqnarray}
with ${\mathcal A}_{11}(I)$ as one matrix element of the matrix for the summed amplitudes as given
in Eq.(13):
\begin{equation}
{\mathcal A}(I) = \left [ I - i \frac{m^2\beta}{(4\pi)^2}{\mathcal M}_{[I]}^{(0)}(E) \right ]^{-1}.
\end{equation}
We will use the above formula for $F_1(q^2) + F_2(q^2)$ to describe Babar results.
\par
We will use our theoretical results to perform a fit by combining the BES data, Babar data in \cite{Bar1} and
the data of the cross section of $p\bar p$ elastic scattering in \cite{PDG}. For the BES data we use the measured
results with BES3 experiment\cite{NBES}. In the fit we are able to determine the coupling constants in the effective theory,
i.e., $c_{0,1}$ and $d_{0,1}$. These constants are complex as discussed before. We make two fits with different theoretical
results in our approach. One fit is done with tree-level results, another is performed with
$N\bar N$ scattering amplitudes in which part of partial wave amplitudes includes summed high-order corrections as indicated
in Eq.(9-13) and in Eq.(15,18).
The tree-level results for $p\bar p$ elastic scattering can be
obtained from the effective theory in Eq.(3). In the fits we start to gradually add more data points
to those nearest to the threshold
until the $\chi^2/d.o.f$ larger than $1$. We do not take these data points below the $n\bar n$-threshold into account
because we assume the isospin symmetry. Our fitting results are shown in Fig.4 for BES and Babar
and in Fig.5 for the cross section.
\par
\begin{figure}
\centering
\includegraphics[width=0.45\textwidth]{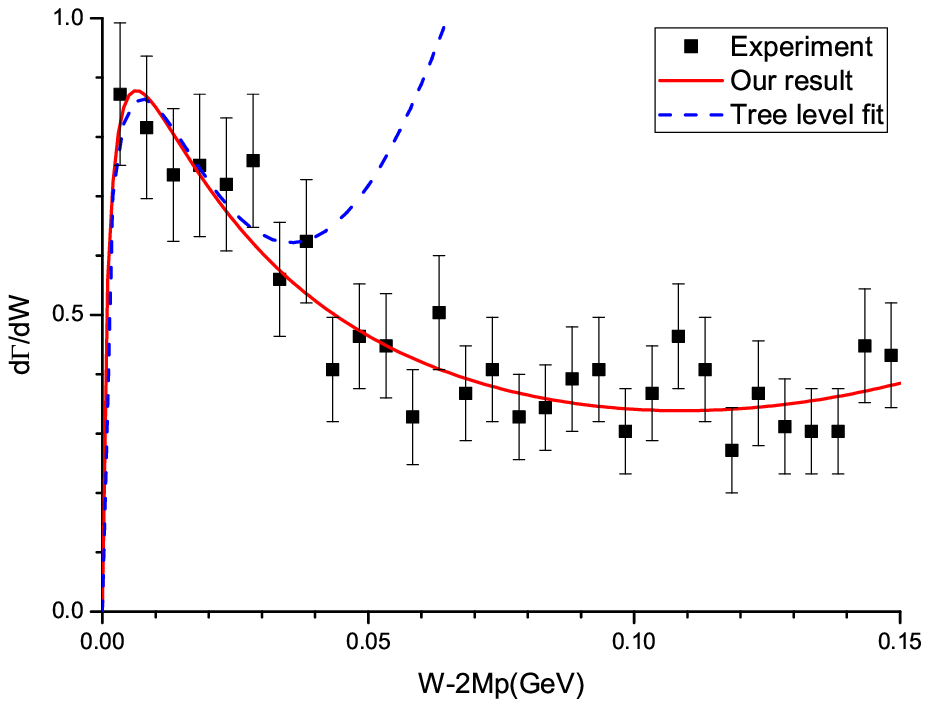}
\hspace{0.2cm}
\includegraphics[width=0.45\textwidth]{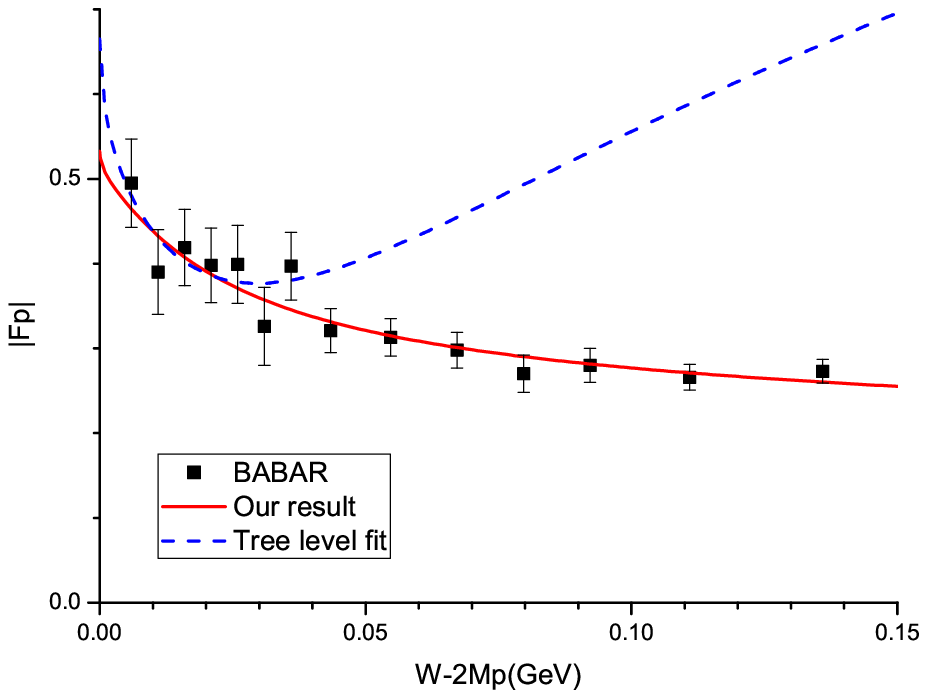}%
\caption{The fitting results for the enhancement observed in $J/\psi \to \gamma p\bar p$ observed by BES(left plot)
and in $e^+e^-\to p\bar p$ by Babar(right plot). $W$ is the total energy of the $p\bar p$ system. }
\label{methodA}
\end{figure}
\par
For the fit with tree-level results we are only able to describe the experimental data in the region with $E <\sim 20{\rm MeV}$.
This can be seen from Fig.4 and Fig.5. Although the $\chi^2/d.o.f$ of the fit for experimental data in this region
is around $1$, but the coupling constants are determined with errors from $40\%$ to $100\%$ or even more.
With the results containing the summed or rescattering partial waves the regions of data which can be fitted become
larger as shown in Fig.4 and Fig.5. For BES data and Babar data the region is with $E\leq 150{\rm MeV}$. For the cross section
of the $p\bar p$ elastic scattering the region is with $E\leq 70{\rm MeV}$. The results of the fit for the coupling
constants in unit of ${\rm GeV}^{-2}$ are:
\begin{eqnarray}
 c_0 = 123(15)-22(11)i, \ \  c_1 = -177(55)-57(27)i,\ \
 d_0=32(31)+17(60)i, \ \  d_1 =73(28) -13(19)i.
\end{eqnarray}
The other constants appearing in $J/\psi\to \gamma p\bar p$ and $e^+ e^-\to p\bar p$ and $\chi^2/d.o.f.$ from
the fit are:
\begin{eqnarray}
&& {\mathcal A}_{p\bar p} +{\mathcal A}_{n\bar n} = 0.23(37), \ \ \ {\mathcal A}_{p\bar p} -{\mathcal A}_{n\bar n} = 2.22(72),
\nonumber\\
&&\tilde G_M^{(p)} + \tilde G_M^{(n)} = 1.18(47), \ \ \ \tilde G_M^{(p)} - \tilde G_M^{(n)} =-0.11(20),
\nonumber\\
&& \chi^2/d.o.f. = 50/63.
\end{eqnarray}
In the above the coefficients ${\mathcal A}_{p\bar p,n\bar n}$ are arbitrarily normalized.
From the fitting results the coupling $d_0$ is not well determined. From Fig.5 one can also see that without these
contact terms in the effective theory the $p\bar p$ elastic scattering can not be described
with pion exchanges. This is also true if we include rescattering effects in $S$-waves.
This is expected as discussed at beginning because without these contact terms
in our effective theory the effects
of a $N\bar N$ annihilation into virtual- and real pions are not included.
These effects are important for $N\bar N$ scattering and should be not neglected.
In our effective theory, the annihilation near threshold is described with these contact terms.
\par
\begin{figure}
\centering
\includegraphics[width=0.45\textwidth]{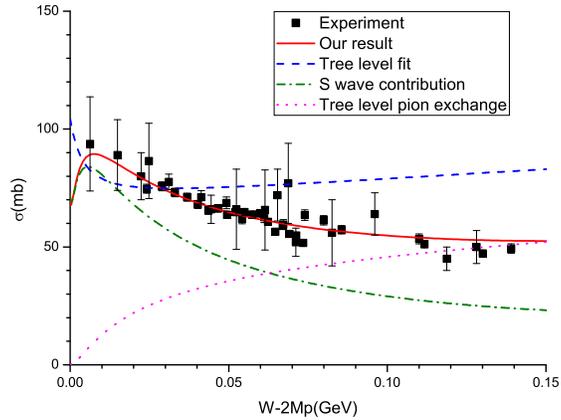}
\caption{The fitting result for the cross-section of the elastic $p\bar p$-scattering near the threshold}
\label{meB}
\end{figure}
\par
From Fig.5 we can see that in our approach for the $p\bar p$ elastic scattering near the threshold
the $S$-wave amplitudes are dominant. We can determine the scattering lengthes of $S$-waves.
In our notation the phase-shift and scattering length of a partial wave is given as:
\begin{equation}
 \exp(2 i \delta_{[j\ell\ell's,I]}(E)) -1 = i \frac{mp}{ 8\pi^2} {\mathcal T}_{[j\ell\ell's,I]} (E), \ \ \
   p\cot \delta_{[j\ell\ell's,I]}(E)) = -\frac{1}{a_{[j\ell\ell's,I]}} + {\mathcal O}(p^2).
\end{equation}
From our fitting results we have the $S$-wave scattering lengthes in unit of ${\rm fm}$
\begin{eqnarray}
&& a_{[0000,0]} = -1.80(19) + 0.32(17) i, \ \ \  a_{[0000,1]} = 2.61(81) + 0.84(41) i,
\nonumber\\
&& a_{[1001,0]} = -0.48(46)- 0.25(89) i, \ \ \  a_{[1001,1]} = -1.07(41) + 0.19(29) i.
\end{eqnarray}
These results are comparable with LEAR experiment\cite{Tbes2} and model results\cite{Tbes3}.
The scattering lengthes are determined by the coupling constants in Eq.(3,9).
We notice here that our numerical results of $c_{0,1}$ in Eq.(20) are roughly twice larger than the corresponding
coupling constants in the effective theory of a $NN$ system\cite{EFTN}. But the determined scattering lengthes
here are much smaller than the corresponding scattering lengthes of the $NN$ systems, because
the relation between scattering lengthes and coupling constants is different in the two different effective theories.
\par
Before summarizing our study we briefly discuss our prediction relevant for the enhancement
in $J/\psi\to \gamma p\bar p$. The measured spectrum can be fitted with the decay amplitude
as an $S$-wave Breit-Wigner resonance form. The fitting done by BES gives the resonance mass
around $1861$MeV and the width $\Gamma <38$MeV. The mass is near and below the threshold.
We have observed that our formula in Eq.(15)
can re-produce the shape of the $S$-wave Breit-Wigner resonance form above the threshold only by tuning
the parameters ${\mathcal A}_{p\bar p,n\bar n}$ . If there is a resonance or structure below and near the threshold,
one should also have roughly the same shape of the $S$-wave Breit-Wigner resonance below the threshold.
But, our amplitude in Eq.(15) below the threshold through analytical continuation
has a cut near the threshold because of the $L_y$-function in Eq.(8) appearing in the amplitude.
Therefore, we can not conclude from our result that there exists  a resonance or structure below the threshold,
although our result agrees with experimental data above the threshold.
\par
To summarize: We have proposed an effective theory to study $N\bar N$ scattering near the threshold.
Because scattering lengthes of $N\bar N$ system are rather small, we have implemented the standard
minimal subtraction scheme to the effective theory to establish a power counting. The power counting
is used to determine the relative importance of higher order corrections. We have found that
certain higher order corrections represented as multiple rescattering can be simply summed. Using
these rescattering amplitudes and assuming that the enhancement
in $J/\psi\to \gamma p\bar p$ and $e^+e^-\to p\bar p$ near the threshold of the $p\bar p$ system
is due to final state interactions,
we can simultaneously explain the experimental data of the enhancement and of the cross section
of $p\bar p$ elastic scattering near the threshold. The $S$-wave scattering lengthes are determined
which are comparable with existing results. Given the fact of the successful description
of experimental data, the proposed effective theory needs to be studied in more detail.
In the future
we will study higher order corrections from loops formed through pion exchanges and the renormalization
group of coupling constants in the effective theory.
In this letter we have ignored the one-loop corrections to dispersive parts of scattering amplitudes,
especially, the one-loop dispersive parts formed through pion exchanges. It has been found in $NN$-scattering
that these parts can receive large corrections in \cite{EFTN1}. It will be interesting to see
if this also happens in $N\bar N$ scattering. This question can only be answered after our undergoing
study of the complete one-loop correction has been done.

\par\vskip20pt

\vskip 5mm
\par\noindent
{\bf\large Acknowledgments}
\par
We would like to thank Prof. H.Q. Zheng for interesting discussions.
This work is supported by National Nature Science Foundation of P.R. China(No. 10721063,10575126 and 10975169).
\par

\end{document}